# Failure modes of irregular ceramic foam under compression: Development of a new image based strut segmentation strategy

Vinit Vijay Deshpande[1] and Romana Piat[1*]

[1] Department of Mathematics and Natural Sciences, University of Applied Sciences Darmstadt, Schöfferstraße 3, Darmstadt 64295, Germany

* Corresponding author: romana.piat@h-da.de

**Abstract:** The work investigates the failure modes of microstructure of an irregular ceramic foam subjected to uniaxial compression loading. The foam material is manufactured using direct foaming method and has polydispersed pores homogeneously distributed in the microstructure. The effective stress-strain curve and the macroscopic strength of the material differs significantly with the predictions of the Gibson-Ashby model that assumes regular microstructure. The objective of this work is to determine the reasons. The work builds upon an image segmentation algorithm that utilizes skeletonization method followed by geometric pruning strategies to isolate the struts in the foam microstructure. In this work, a novel pruning strategy defined by a physics-based significance measure is proposed which identifies the struts whose failure leads to macroscopic failure of the material. The stresses in the struts are calculated by a finite element simulation which are then utilized to determine their failure modes. This also reveals the relationship between the struts' orientation and their failure modes. The energy dissipated by the individual failure modes as the loading is increased shows that there are two dominant modes which are different from the bending failure reported in the Gibson-Ashby model. These failure modes are mixed compression-tension and a pure compression.

Keywords: Failure modes, microstructure segmentation, irregular foam, strut, compression strength

## 1. Introduction

Open cell ceramic foams are a special type of cellular materials composed of interconnecting network of slender structures called struts. The base material of these struts is ceramic which makes them resistant to high temperature and high corrosive environments. This along with high porosity makes the overall foam material highly sought after in applications like molten metal filtration, exhaust gas filters, biomedical implants and light weight construction [1]. The traditional methods of manufacturing ceramic foams can be classified into two types. One in which the pores are formed indirectly by mechanically stirring a ceramic solution e.g. direct foaming [2], gel casting [3], etc. The second in which the shape and size of pores are directly controlled by replica technique [4], 3D printing [5] among others. The first method is simpler in execution and scalable but fails to control the pore size in the microstructure leading to non-uniform size distribution of the pores. The second method has precise control on pores sizes but is expensive, requires precise control of process parameters and is difficult to scale.

A critical parameter of ceramic foam is its compression strength. It is subjected to compressive loading in many applications [6] and the highly porous nature means that it is susceptible to cracking. Hence to maintain the structural integrity, it is extremely important to study this property. The most influential theoretical works concerning effective material properties of foams were carried out by L.J Gibson, M. F. Ashby and colleagues in [7, 8]. They developed proportionality laws relating effective plastic collapse stress (for metallic foams) and effective crushing strength (for brittle foams) with the porosity. The laws are based upon the idea that the effective response of the foam is related to the response of the individual struts in the foam. For brittle foams, the effective uniaxial compression strength is related to bending strength of the struts. Assuming that all the struts in the foam have identical dimensions, the section modulus of the strut is expressed in terms of the volume fraction which lead to proportionality law between crushing strength (compression strength) of the foam and fracture strength of the base ceramic material. The proportionality constant is determined from experimental measurements.

Experimental studies on measurement of uniaxial compression strength of brittle foams were performed in [6, 9]. An interesting observation in these studies was that the effective stress-strain curve of the foam material depends on the nature of the boundary conditions. If the loading platens are firmly attached to both sides of the foam in such a way that the layers of cells near the loading platens are prevented from failing, then the effective stress increases with the load till there is an abrupt failure. If one loading platen that applies the load is just resting on the foam material, then the maximum stress value is much lower, and the row of cells touching the loading platen fails first followed by the second row and so on. Even with this difference, the microstructure mode of failure in both cases is strut bending as defined in the Gibson-Ashby (GA) models.

The uniaxial compression behaviour of cellular architected materials has been studied extensively to understand deformation modes of individual struts [10]. Since these materials are manufactured using 3D printing, they have highly regular microstructure with struts of uniform size. [10] showed that the deformation modes of struts are either stretching dominated or bending dominated depending on the how many struts are attached at the junction point. At low volume fractions, stretching dominated structures showed higher stiffness and strength than the bending dominated structures [11]. Three types of cellular architectures namely kelvin, cuboctahedron and octet were studied in [12] through experiments and simulations to understand the strut failure modes. An important observation was that the stress concentration at the junction points played more important role in determining the macroscopic compression strength than the individual strut failure modes. It was also observed that in kelvin architecture, as the volume fraction is increased, the compressive stresses in the struts delay the strut failure in bending resulting in higher compressive strength than that predicted by the GA model.

The present paper studies an alumina foam manufactured by direct foaming as reported in [13]. The microstructure has polydisperse spherical shaped pores distributed homogeneously throughout the material space. The detailed description of the manufacturing and the measurement of the effective material properties can be found in [13]. In [14], the same authors performed experimental measurements and computer simulations to determine the effective stress-strain curve and the strength of the foam when subjected to uniaxial compressive loading. It will be shown in the present paper that the strength value of the foam reported in [14] is much higher than that predicted by the GA model. Our recent article [15] performed computational studies to determine the compression failure behaviour of the same material. The results on effective compression stress-strain behaviour and the strength value concurred with the experimental and numerical observations in [14]. The discrepancy between the strength values and the effective stress-strain curves obtained in the above studies with the theoretical works is the inspiration behind our present paper.

The paper answers three questions:

1. Why does the theoretical model of Gibson-Ashby (GA) not work in the case of the studied foam?
2. Why is the stress-strain curve very different than the one reported in the GA models?
3. Are the failure modes in the studied material different than the one assumed in the GA model?

In order to answer these questions, the paper is organised into following sections. Section 2 introduces the material and its uniaxial compression behaviour. Section 3 discusses an image segmentation algorithm and the pruning strategies that isolate the struts in the microstructure. Section 4 presents the size and orientation distribution of the segmented struts. Section 5 describes the different failure modes of the struts by correlating the results of the image segmentation algorithm and the finite element simulations. Section 6 discusses the effect of different failure modes on the effective compression stress-strain behavior of the studied material. Section 7 discusses the results while section 8 concludes the work.

## 2. Study of alumina foam

The microstructure of alumina foam was obtained by micro-CT scanning a material sample. Fig.1a shows a 3D binary image of the foam microstructure. This microstructure is irregular in the sense that it cannot be simplified into a single representative unit cell. The black color in the image represents porosity and the white color is alumina material. The volume fraction of the alumina in the foam is 0.255. Three cylindrical shaped samples of the same microstructure were manufactured and tested in [14] to determine the effective uniaxial compression stress-strain behaviour. The size of this material sample shown in Fig.1a is too big to convert it into a finite element model. Therefore, a statistical study was performed in [15] to determine the appropriate size of the material sample that can be used for finite element modelling. The FE model of this volume element (VE) along with the boundary conditions can be seen in Fig.1b. The edge length of this VE is 25% that of the foam sample in Fig.1a.

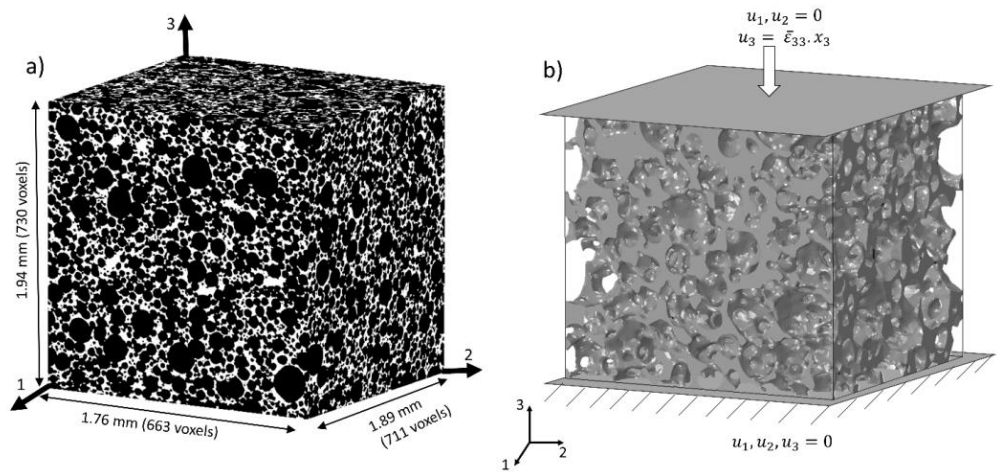

**Fig.1.** *a) 3D binary image of microstructure of alumina foam; b) material volume element and boundary condition for uniaxial compression simulation.*

## 2.1 Material modelling

A surface mesh with triangular elements is generated from the 3D binary image in MATLAB [16] which is then further converted into a volume mesh in ABAQUS [17]. Linear tetrahedral elements are chosen for the purpose of mesh creation. The constitutive behavior of the alumina base material is modelled using Johnson Holmquist 2 softening plasticity model [18] whose yield function f is a function of stress tensor $\boldsymbol{\sigma}$ and a scalar damage variable $D$ that takes any value between 0 and 1:

$$f(\boldsymbol{\sigma}, D) = \sigma_{eq}(\boldsymbol{\sigma}) - \sigma_y(\boldsymbol{\sigma}, D), \tag{1}$$

where, $\sigma_y$ is the yield stress, $\sigma_{eq}(\boldsymbol{\sigma}) = \sqrt{\frac{3}{2}\boldsymbol{s}:\boldsymbol{s}}$ is the von Mises stress such that $\boldsymbol{s} = \boldsymbol{\sigma} - p\boldsymbol{I}$ is a deviatoric stress tensor and $p$ is the pressure. The yield stress, $\sigma_y^*(\boldsymbol{\sigma}, D)$ for the JH-2 model is defined as,

$$\sigma_y^*(\boldsymbol{\sigma}, D) = (1 - D)\sigma_i^*(\boldsymbol{\sigma}) + D\sigma_f^*(\boldsymbol{\sigma}). \tag{2}$$

The superscript * indicates that the stress value has been normalized with respect to Hugoniot elastic limit $\sigma_{HEL}$. The subscripts $i$ and $f$ refers to intact and fractured material strengths

$$\sigma_i^*(\boldsymbol{\sigma}) = A\left(\frac{T + p(\boldsymbol{\sigma})}{P_{HEL}}\right)^n (1 + C \ln \dot{\bar{\varepsilon}}_p^*), \tag{3}$$

$$\sigma_f^*(\boldsymbol{\sigma}) = B\left(\frac{p(\boldsymbol{\sigma})}{P_{HEL}}\right)^m (1 + C \ln \dot{\bar{\varepsilon}}_p^*). \tag{4}$$

The intact and failed strength as a function of pressure $p$ can be seen in Fig. 2a. $T$ is hydrostatic tensile strength. The rate dependency of the material is scaled by the parameter $C$. $\dot{\bar{\varepsilon}}_p^* = \frac{\dot{\bar{\varepsilon}}_p}{\dot{\bar{\varepsilon}}_p^0}$ is the normalized equivalent plastic strain rate where $\dot{\bar{\varepsilon}}_p^0$ is a reference strain rate below which $C = 0$. The equivalent plastic strain rate is $\dot{\bar{\varepsilon}}_p = \sqrt{\frac{3}{2}\dot{\boldsymbol{e}}_p : \dot{\boldsymbol{e}}_p}$ where $\dot{\boldsymbol{e}}_p$ is deviatoric plastic strain rate. $A, B, C, m, n$ are material constants and $P_{HEL}$ is pressure at Hugoniot elastic limit.

The damage $D$ is defined in terms of accumulation of equivalent plastic strain such that,

$$D = \sum_{i=1}^{st} \frac{\Delta \bar{\varepsilon}_{p_i}}{\bar{\varepsilon}_p^{max}(\sigma)}, \tag{5}$$

where, $\Delta \bar{\varepsilon}_{p_i}$ is increment in equivalent plastic strain at integration step $i$ and the summation indicates summing all the strain increments from start of loading till the current step $st$. $\bar{\varepsilon}_p^{max}(\sigma)$ is equivalent plastic strain to fracture at constant pressure $p(\sigma)$ given by,

$$\bar{\varepsilon}_p^{max(\sigma)} = d_1 \left( \frac{T + p(\sigma)}{P_{HEL}} \right)^{d_2}. \tag{6}$$

This relationship is shown in Fig. 2b. $d_1$ and $d_2$ are material parameters. The plastic flow rule is given by,

$$\dot{\varepsilon}_p = \dot{\lambda} \frac{\partial g}{\partial \boldsymbol{\sigma}}, \tag{7}$$

where the plastic potential function is $g = \sigma_{eq}(\boldsymbol{\sigma}) = \sqrt{\frac{3}{2} \boldsymbol{s} : \boldsymbol{s}}$.

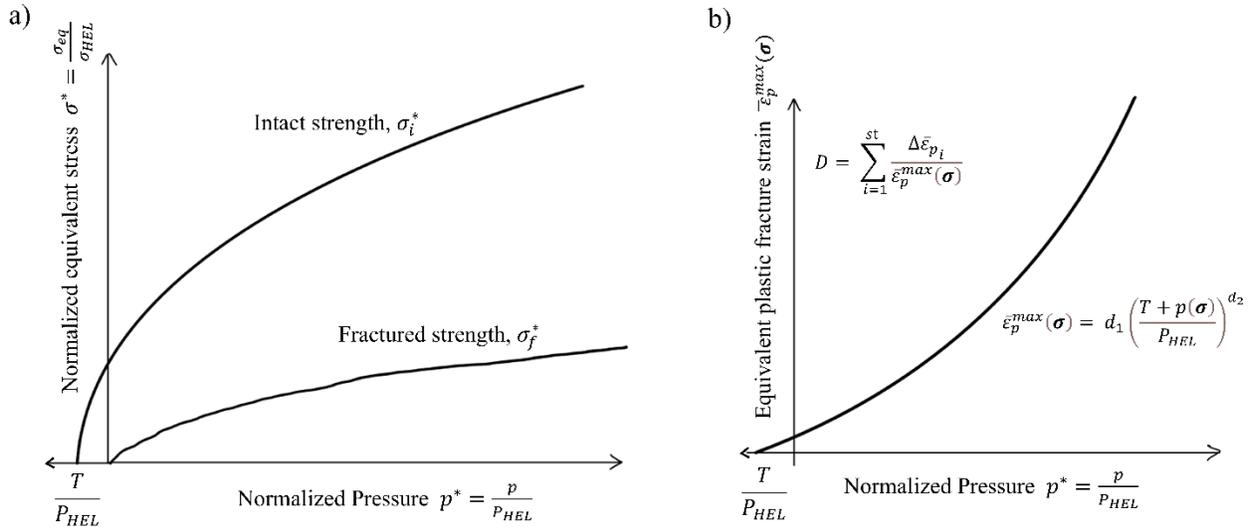

**Fig. 2.** *a) Strength model and b) damage model of Johnson Holmquist 2 model.*

The equation of state before damage (i.e. $D = 0$) relating hydrostatic pressure with volumetric strain is given by,

$$p = K_1 \mu + K_2 \mu^2 + K_3 \mu^3 \quad (\mu > 0), \tag{8}$$

$$p = K_1\mu \ (\mu < 0),$$

where, $\mu$ is the volumetric strain. After damage begins to accumulate, bulking occurs in the material which is leads to increase in the pressure. Therefore, the equation of state becomes,

$$p = K_1\mu + K_2\mu^2 + K_3\mu^3 + \Delta p . \tag{9}$$

$K_1, K_2, K_3$ are material constants. $\Delta p$ is calculated by equating the elastic energy loss during damage to the potential hydrostatic energy.

The material constants for the 99.5 % alumina which was used as a base material of the foam are directly adopted from [19]. Table 1 shows the values of the material constants.

| Material parameters | Notation | Value |
|---|---|---|
| Density | $\rho_0$ (kg/m$^3$) | 3890 |
| Shear modulus | G (GPa) | 152 |
| Intact strength coefficient | A | 0.88 |
| Intact strength exponent | N | 0.64 |
| Fractured strength coefficient | B | 0.28 |
| Fractured strength exponent | M | 0.6 |
| Rate dependency coefficient | C | 0 |
| Hydrostatic tensile strength | T (MPa) | 150 |
| Normalized maximum fractured strength | $\sigma_{max}^f$ | 0.2 |
| Hugonoit elastic limit | HEL (GPa) | 6.57 |
| Pressure at Hugonoit elastic limit | $P_{HEL}$ (GPa) | 1.46 |
| Pressure constants | $K_1$ | 231 |
| | $K_2$ | -160 |
| | $K_3$ | 2774 |
| Damage coefficient | $d_1$ | 0.01 |
| | $d_2$ | 0.7 |

**Table 1.** *Material parameters of 99.5 % alumina for Johnson Holmquist 2 model.*

Fig. 3a shows effective compression stress-strain curve obtained from the displacement and the reaction force calculated at the top face of the VE [15]. The simulation is performed along three orthogonal directions as indicated in the grey color curves in the figure. It can be seen that the stress increases till it reaches a peak stress value after which there is a gradual loss of strength till there is complete failure. The three curves in black color are the experimentally measured curves on three different specimens as described in [20]. The experimental data was measured only till the point at which the maximum stress is reached. The average maximum stress from the three simulation curves is 65.2 MPa while that from the three experimental curves is 65.7 MPa. According to the simulation data, the maximum compressive stress is reached at the strain value of 0.33%. Fig.3b shows the damaged regions in the VE at the last step of the loading. These damaged regions indicate damaged struts which are distributed homogeneously throughout the microstructure.

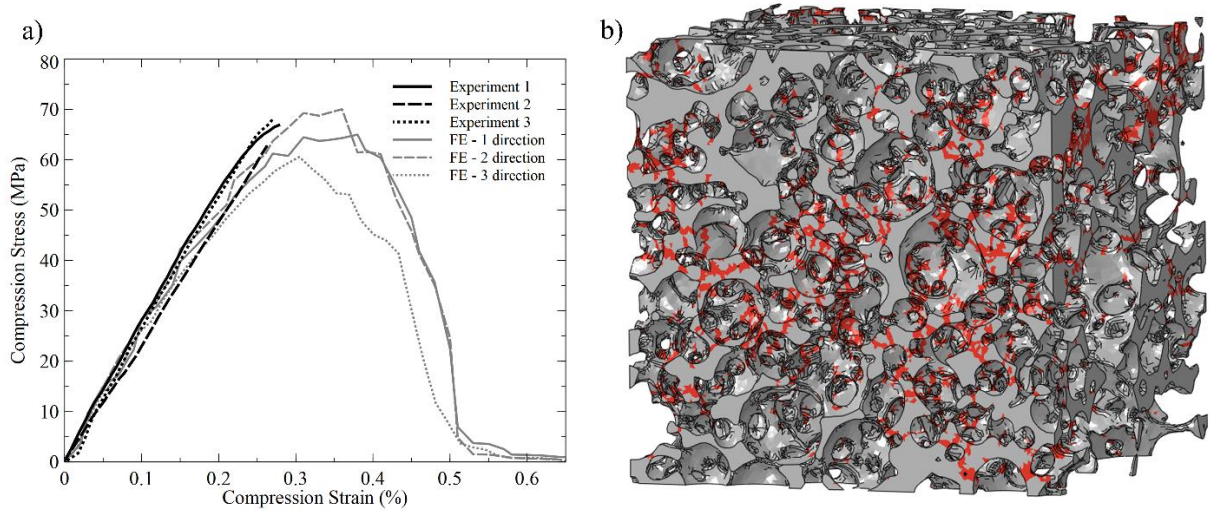

**Fig. 3.** *a) Effective compression stress-strain curves obtained from simulation along three orthogonal directions [15] and experimentally measured curves on three samples [11]; b) VE with red color indicating damaged regions at the last step of the loading.*

## 2.2 Theoretical model of compression strength versus relative density

Maiti, Gibson and Ashby in their seminal work in [7] developed analytical expressions that relate the mechanical response of cellular solids to that of the individual struts. It assumes a regular microstructure composed of cells with identical dimensions. A unit cell of the foam material is shown in Fig.4b. It is a cuboidal unit cell with struts of identical dimensions. The theory states that a remotely applied uniaxial loading, $\sigma_{cr}$ on the foam structure (Fig.4a) leads to a loading **F** on individual struts (Fig.4b-c) which in turn causes bending failure in them. This bending failure occurs when the maximum tensile stress at the surface of the strut equals modulus of rupture of the strut material. The expression relating crushing strength, $\sigma_{cr}$ of the foam to the modulus of rupture $\sigma_{fs}$ of base material is shown in Eq.10.

$$\sigma_{cr} = C\sigma_{fs} \left(\frac{\rho_{cellular}}{\rho_{base}}\right)^{\frac{3}{2}}, \quad (10)$$

where, $\rho_{cellular}$ and $\rho_{base}$ indicate density of cellular solid and base material respectively. Eq.10 correctly captures the critical stress of the foam structure whose effective stress-strain curve takes the form shown in Fig. 4d as reported in [7, 21]. In this figure, after the initial linear elastic region, there is a sudden drop in the stress value because of bending failure of the struts in the top row cells of the material sample that are in immediate contact with the loading platen. The subsequent undulations in the curve are the result of failure of the struts in the subsequent rows of cells. Since the mode of failure of all the struts is same and the struts have identical dimensions, the mean stress value remains constant in this region (plateau). At higher loadings, there is densification of the material which causes steep rise in the stress value.

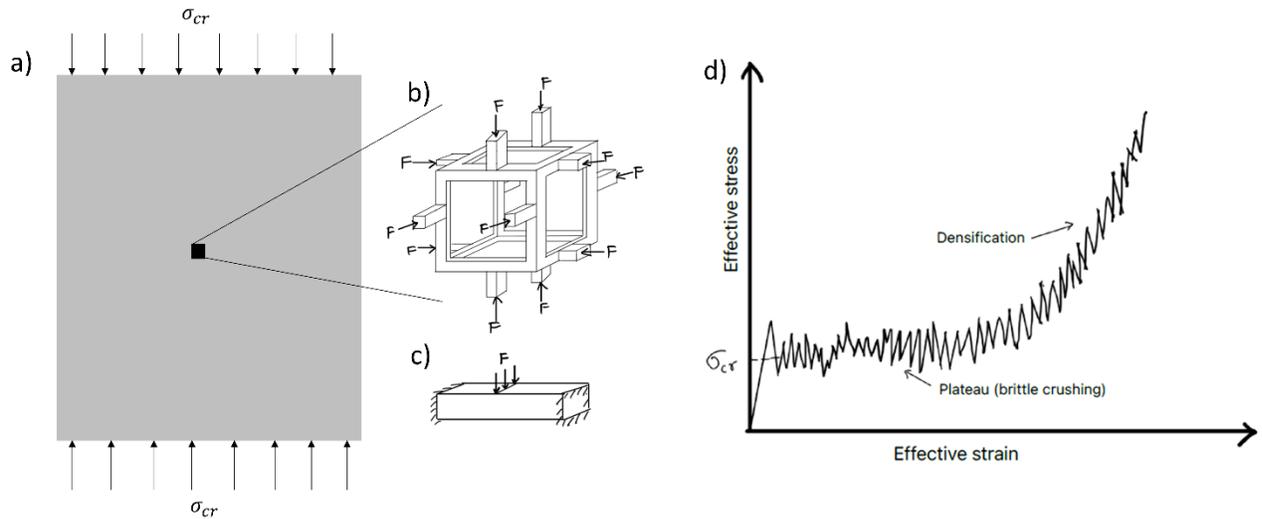

**Fig. 4.** *a) Foam structure subjected to remote loading $\sigma_{cr}$; b) an idealized unit cell of the foam; c) force **F** on the strut; d) typical effective stress-strain curve of an open cell brittle foam as reported in [21].*

Direct comparison between Fig.3a and Fig.4d shows that the nature of stress-strain curves is completely different. The value of peak stress in Fig.3a is around 65MPa whereas that predicted by Eq. 10 is 11.58 MPa considering tensile strength of alumina as 450MPa [19], relative density of foam is 25.5% and value of the constant $C$ is 0.2 [21].

The objective of the present paper is to investigate the reasons behind this difference. It will be shown in the subsequent sections that this is because of the difference in the microstructure and the failure modes of the struts.

## 3. Segmentation of struts

The weakest members in the microstructure are the struts and hence the macroscopic failure of the material occurs when the struts lose their load bearing capacity. Studying how and when the struts fail with respect to the macroscopic loading can reveal the correlation between strut failure modes and the macroscopic failure. This requires identifying each strut in the microstructure image and then studying its geometry and its response to loading. It can be seen in Fig.1 that the microstructure of the studied material is quite complex with struts having a wide range of distribution in their shape, size and orientation. Therefore, identifying or segmenting these struts in the microstructure is not a trivial problem. Typical methods for segmenting slender features in the microstructure e.g. particles or fibres in composites utilize filtering methods that have a fixed size and shape of a kernel [22]. Examples include anisotropic Gaussian filtering, Hessian matrix and structure tensor. These methods work very well when the features have the same size and shape throughout the microstructure. However, their accuracy reduces drastically in the presence of non-uniform features. In our recently published paper [23], a generalized image segmentation algorithm was developed which identifies the struts in the microstructure on the basis of their geometric properties. A novel pruning strategy based on cross-sectional area of the strut was developed that was able to isolate the struts irrespective of their size and shape. In the present article, this algorithm has been developed further to identify specific struts in the microstructure whose failure led to the macroscopic failure of the material. This section describes the image segmentation algorithm proposed in [23] in brief followed by a detailed description of how it is further developed to determine the strut failure modes. Fig. 5 explains the steps involved in the segmentation with the grey color shaded regions indicate contributions of the present article.

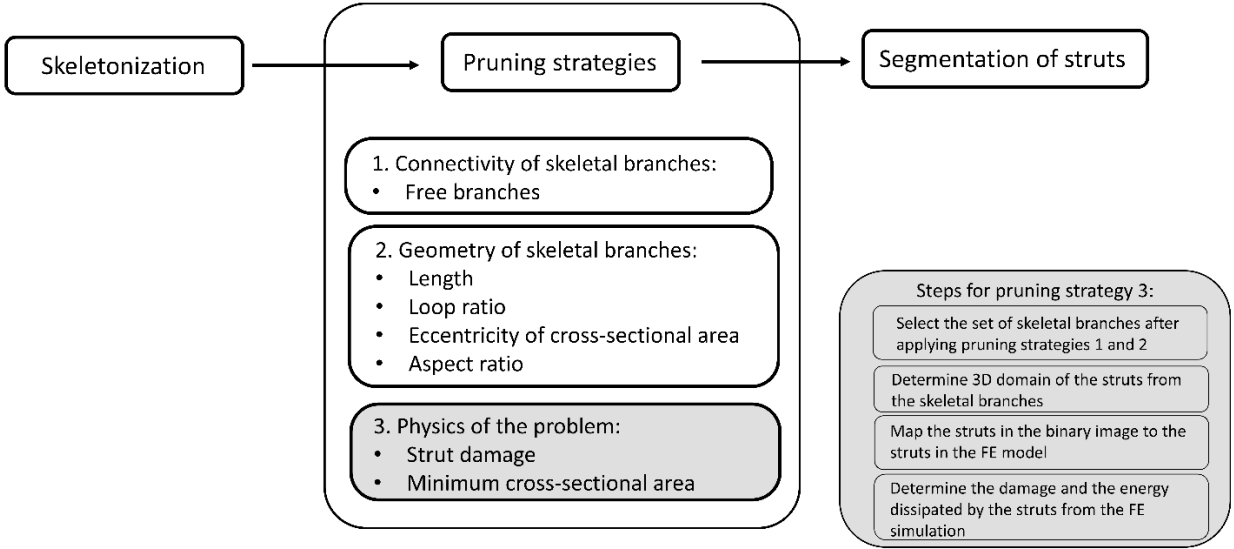

**Fig. 5.** *Image segmentation procedure for segmentation of struts in the microstructure.*

### 3.1 Skeletonization

The first step is the conversion of the foam microstructure volume element shown in Fig.1b into a single voxel thick skeleton. The skeleton is calculated in such a way that the topology of the microstructure defined by the Euler characteristic does not change during skeletonization. The Euler characteristic of a 6-connected foreground geometry in a binary image is defined as,

$$G_6 := \#O - \#H + \#C, \tag{10}$$

where a 6-connected geometry in a digital image is a set of voxel points of the same value in the 6-neighbourhood of each other. Detailed definitions can be found in the [23]. The skeleton can be used to study the orientation and the thickness of the struts in the microstructure. However, skeletonization algorithms are highly susceptible to image noise and insufficient resolution. It creates a lot of artificial skeletal branches that do not represent the actual geometry of the microstructure. Removal or pruning of these artificial branches is done by defining pruning strategies. A brief description of these strategies is given below.

### 3.2 Pruning strategies

The strategies described in sections 3.2.1 and 3.2.2 are based on the interconnectivity of the skeletal branches and their geometry respectively. They were proposed in our article [23].

### 3.2.1 Pruning based on connectivity (pruning free branches)

Any skeletal branch that represents the actual geometry of the microstructure is always connected to other skeletal branches at both its ends. An exception to this case is the skeletal branches that represent microstructure features that lie at the boundary of the volume element. These features are literally cut at the boundary faces and hence their skeletal branches also have a free end lying at the boundary face. These skeletal branches represent an actual feature of the microstructure and hence are not artificial branches. Moreover, during the macroscopic loading of the VE, these boundary features might get deformed and possibility damaged depending on the boundary conditions. Hence, it is necessary to keep these branches. Hence in this pruning strategy, any skeletal branch that lie entirely in the interior of the VE and has one free end is pruned. An illustrative example is shown in Fig. 6.

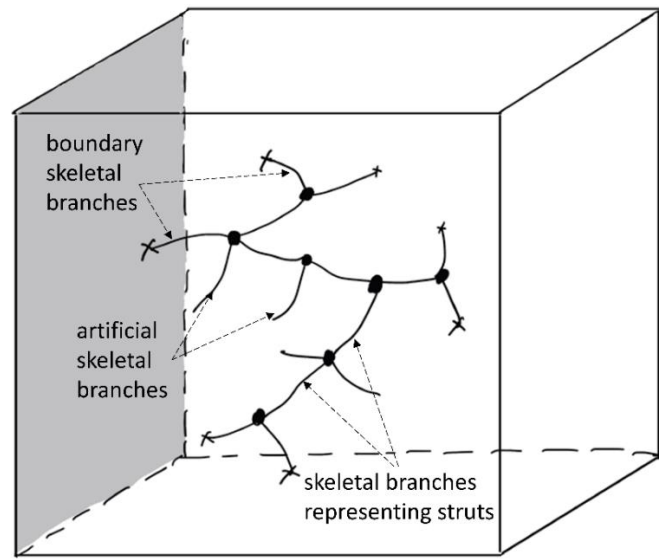

**Fig. 6.** *A volume element showing skeletal branches that represent actual struts, boundary skeletal branches with one end touching the boundary being cross-marked and artificial branches (one free end).*

### 3.2.2 Pruning based on geometric significance measures

#### 3.2.2.1 Length of the skeletal branch

The length of the $i^{th}$ skeletal branch is defined as the total number of voxels present in it. If T($i$) is a set of voxels that defines the $i^{th}$ skeletal branch, then its length $l_i$ is defined by the cardinality of this set, i.e. $l_i := \#(T(i))$. An $i^{th}$ skeletal branch is pruned if its length is less than or equal to a threshold parameter $\lambda_{length}$, i.e. $l_i \leq \lambda_{length}$.

### 3.2.2.2 Loop ratio of the skeletal branch

Loop Ratio LR($i$) of the $i^{th}$ skeletal branch is the ratio of Euclidean distance between the end points of a skeletal branch to its length., i.e. $\text{LR}(i) := \left\| r_{q_1} - r_{q_{n_i}} \right\| / l_i$ where $q_j \in T(i) \mid 1 \leq j \leq n_i$. $r_q$ is the position vector of point q and $n_i$ is the total number of points in skeletal branch T($i$). An $i^{th}$ skeletal branch is pruned if its Loop Ratio is less than or equal to a threshold parameter $\lambda_{LR}$, i.e. $\text{LR}(i) \leq \lambda_{LR}$

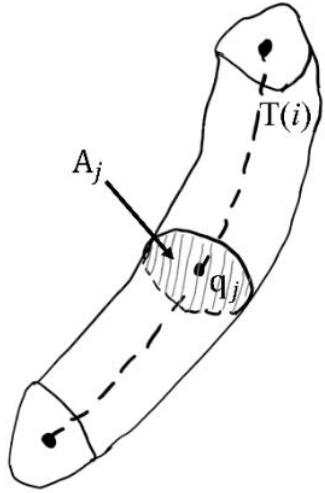

**Fig.7.** $i^{th}$ *skeletal branch* T($i$) *and the geometric feature (strut) that it represents.*

### 3.2.2.3 Eccentricity of the cross-section of the strut

At every voxel point $q_j$ of the $i^{th}$ skeletal branch, the cross-sectional area $A_j$ of the microstructure feature that the branch represents is calculated (refer Fig.7). An equivalent radius $R_j^{eq}$ is defined such that $R_j^{eq} = \sqrt{A_j/\pi}$. Further, eccentricity of the cross-section is defined as $\text{Eccen}_j = R_j^{eq}/(t_j/2)$, where $t_j$ is the thickness of the strut at point $q_j$ obtained from distance transform of the microstructure image. An $i^{th}$ skeletal branch is pruned if maximum eccentricity of the strut is greater than or equal to a threshold parameter $\lambda_{eccen}$, i.e. $\text{Eccen}_{max}(i) \geq \lambda_{eccen}$.

### 3.2.2.4 Aspect ratio of the strut

Aspect ratio $AR(i)$ of the $i^{th}$ struts is defined as ratio of length of the skeletal branch to its minimum thickness, i.e. $AR(i) = \#(T(i))/(t_{min}(i))$. An $i^{th}$ skeletal branch is pruned if the aspect ratio of the strut is less than or equal to a threshold parameter $\lambda_{AR}$, i.e. $AR(i) \leq \lambda_{AR}$.

The above-mentioned pruning strategies are aimed at removing artificial skeletal branches through the selection of appropriate threshold parameters. These parameters are decided by studying the distribution of these geometric quantities and identifying the outliers based on visual observations. The detailed steps in given in [23]. The next pruning strategies are newly introduced in this article that incorporates the physics of the problem to prune the skeletal branches.

### 3.2.3 Pruning based on physics-based significance measures

### 3.2.3.1 Damage in the strut

After applying the above-mentioned pruning strategies, the remaining skeletal branches represent the struts in the microstructure. In this pruning strategy, any strut that does not get damaged during the macroscopic loading is pruned. In order to do that, the strut domain $T^{vol}(i)$ associated with each of the remaining skeletal branch $T(i)$ is identified as described in [23]. Next, the struts in the 3D image are mapped to the corresponding struts in FE model. In order to study the stress state in the struts, the first step is to identify the finite elements that define the strut's volume. Let 'e' be any finite element in element set $\mathbb{EL}$ that represents all the elements of the FE model. Let 'nd' be any node in the node set $\mathbb{Nd}$ that represents all the nodes in the FE model.

$$\mathbb{E} \coloneqq \{e(i) \mid 1 \leq i \leq n_{te}\}, \tag{11}$$

$$\mathbb{Nd} \coloneqq \{nd(i) \mid 1 \leq i \leq n_{tn}\}, \tag{12}$$

$n_{te}$ and $n_{tn}$ are the total number of elements and nodes respectively in the FE model. Each linear tetrahedral element 'e' is made up of 4 nodes, i.e.

$$ndset^e \coloneqq \{nd_1^e, nd_2^e, nd_3^e, nd_4^e\}. \tag{13}$$

Let S be a set that contains all the voxel points that belong to the microstructure (foreground) in the binary image. Let set $\mathbb{F}$ contains all voxel points in S that have the same coordinates as that of the nodes in set $\mathbb{Nd}$, i.e.

$$\mathbb{F} = \mathbb{Nd} \cap S . \tag{14}$$

Let $\mathbb{f}$ be a voxel point in $\mathbb{F}$ such that, $\mathbb{F} = \{\mathbb{f}_i \mid 1 \leq i \leq n_{tn}\}$. Let $F(i)$ be a set of nodes in strut domain $T^{vol}(i)$, i.e.

$$F(i) = \mathbb{F} \cap T^{vol}(i) . \tag{15}$$

Let $E(i)$ be a set of elements in the strut domain $T^{vol}(i)$. These are identified by comparing the nodes of every element 'e' in the FE model with the node set $F(i)$. The elements whose nodes belong to the set $F(i)$ are identified as elements of set $E(i)$.

$$E(i) := \{e \in \mathbb{E} \mid (\text{ndset}^e \cap F(i)) \neq \emptyset\} \tag{16}$$

An example of a strut in shown in Fig. 8a. The set of elements $E(i)$ representing the strut, are shown in Fig. 8b.

During compression loading simulation of the FE model, at each load step 's', the damage in element 'e' is denoted by $D^s$ where $1 \leq s \leq n_{st}$ and $n_{st}$ is the total number of load steps. Note that the damage calculation is explained in detail in section 2.1. A skeletal branch $T(i)$ is deleted if no element in $E(i)$ is damaged for all load steps. i.e.

$$\forall\, e \in E(i), D^s = 0, 1 \leq s \leq n_{st} \Rightarrow \text{the skeletal branch is pruned.} \tag{17}$$

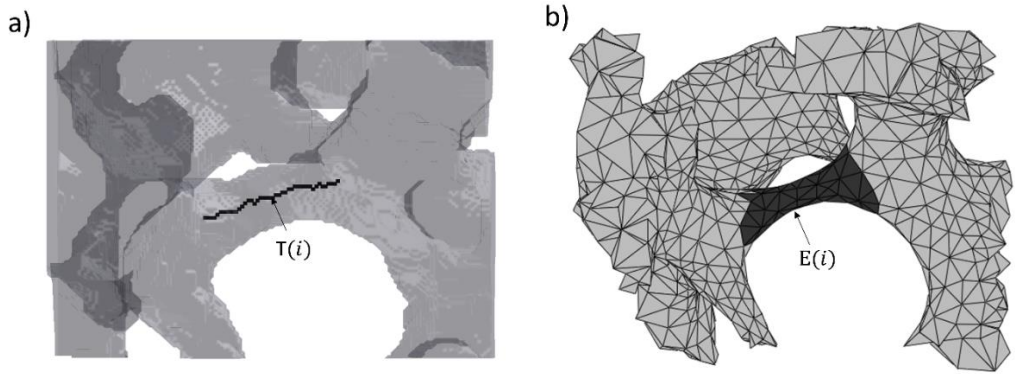

**Fig.8.** *a) A typical skeletal branch $T(i)$; b) finite element set $E(i)$ representing the strut volume.*

### 3.2.3.2 Minimum cross-sectional area of the strut

This pruning strategy identifies and removes the skeletal branches in terms of their minimum cross-sectional area. From the cross-sectional area $A_j$ of every point $q_j$ of the $i^{th}$ skeletal branch, the minimum value $A_{min}(i)$ of the cross-sectional area is calculated. An $i^{th}$ skeletal branch is pruned if $A_{min}(i)$ is greater than a threshold parameter $\lambda_{minA}$, i.e. $A_{min}(i) > \lambda_{minA}$.

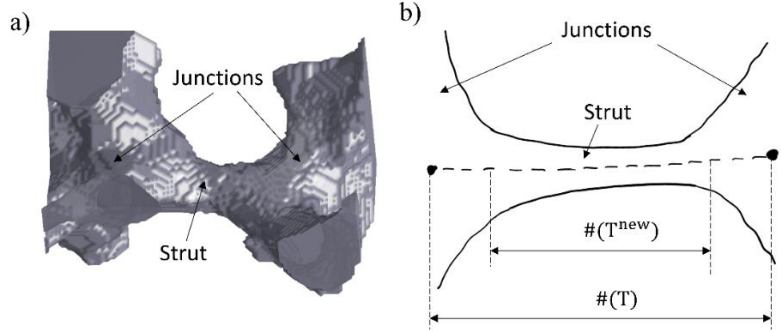

**Fig.9.** *a) An example of an actual strut in the microstructure and b) an illustration showing skeletal branch T(i) and its truncated version $T^{new}(i)$.*

An important observation in Fig. 9a is that a strut is attached to thick junction regions at both the ends. A skeletal branch represents both the strut as well as the junctions (refer Fig. 9b). However, during macroscopic loading, only the thin struts get damaged. The thick junction regions do not lose any load bearing capacity. Therefore, it is necessary to truncate the branch T($i$) into a smaller branch $T^{new}(i)$ that represents just the strut region. This is done by calculating cross-sectional area at all the skeletal points in the $i^{th}$ branch and then selecting only that set of skeletal points that have cross-sectional area smaller a particular threshold as defined in Eq.18. $T^{new}(i)$ is defined in Eq.19.

$$A_j < \lambda_s \cdot A_{min}(i). \qquad (18)$$

$$T^{new}(i) \coloneqq \{q_j \in T(i) \mid A_j < \lambda_s \cdot A_{min}(i)\}. \qquad (19)$$

$\lambda_s$ is a threshold parameter of the truncation operation.

### 3.3.3.3 Threshold parameter for minimum cross-sectional area

The pruning strategy of minimum cross-sectional area was already introduced in our article [23]. The objective of this strategy is not to remove artificial skeletal branches but to remove any branch whose strut is outside of our range of interest. In this work, this range of interest comes from the chosen problem of uniaxial compression failure of the material. It is observed that as the value of the threshold parameter $\lambda_{minA}$ is increased, the skeletal branches representing more and more thicker struts are retained (i.e. not pruned). So, the correct value of $\lambda_{minA}$ is the one in which all the struts that play a critical role in the macroscopic failure of the volume element are captured. In order to achieve this, the following tuning study is performed.

As the VE is compressed, its elastic energy increases. As the elements in the VE get damaged, this elastic energy gets dissipated. Since the objective is to study failure modes, it is important to capture all the struts that dissipate the stored energy. At any load step 's', the energy dissipated $\Psi_d$ by all the captured struts is given by,

$$\Psi_d = \int_s \int_{V_{E_T}} \boldsymbol{\sigma} : \dot{\boldsymbol{\varepsilon}}^{pl} \, dV_{E_T} \, ds, \quad \text{where } E_T = \bigcup_i E(i), \ 1 \leq i \leq \text{total struts captured} \quad (20)$$

$V_{E_T}$ is the volume of $E_T$, $\boldsymbol{\sigma}$ is the Cauchy stress and $\dot{\boldsymbol{\varepsilon}}^{pl}$ is the plastic strain rate. The term $\Psi_d$ is sensitive to $\lambda_{minA}$ as the larger the value of $\lambda_{minA}$, the greater number of struts are captured and hence higher is the value of $\Psi_d$. Thus increasing $\lambda_{minA}$ increases the chances that all the struts that dissipate energy are captured. In the tuning study, three values of $\lambda_{minA}$ namely 1100, 3000 and 4000 (number of voxels) are considered and the graph of $\Psi_s$ vs macroscopic strain is plotted in Fig. 10a. Total energy dissipated by the FE model is also plotted. It can be seen that the curve for $\lambda_{minA} = 4000$ follows the curve of the total dissipated energy till the strain value of 0.5% above which there is some deviation. It can be seen in Fig.3a that the VE losses its strength completely at the strain value of 0.5% so what happens after that is not relevant. Increasing the value of $\lambda_{minA}$ further did not lead to any improvement. Hence, this value of $\lambda_{minA}$ is selected as an appropriate threshold value for this pruning strategy. Note that the method of determining the threshold parameter $\lambda_{minA}$ makes this pruning strategy a physics based rather than a purely geometric one. This strategy is different from the one based on damage because the damage-based pruning removes all the struts that remain undamaged during the loading. The cross-sectional area-based strategy removes the struts that get damaged but do not lead to the loss of macroscopic load bearing ability of the material.

It is observed that in some struts, a small number of elements got damaged but before the strut could fully fail, its neighbouring struts got damaged and the load on the strut in question was released.

Such struts even though technically damaged still have load carry capacity and hence do not contribute to failure mode identification. Removal of such struts from consideration reduced the energy dissipated during the later stages of the loading as shown in the grey color curve in Fig. 10b. This is acceptable since the VE is totally damaged by then. This grey color curve will henceforth be used to indicate total energy dissipated by the VE during the loading.

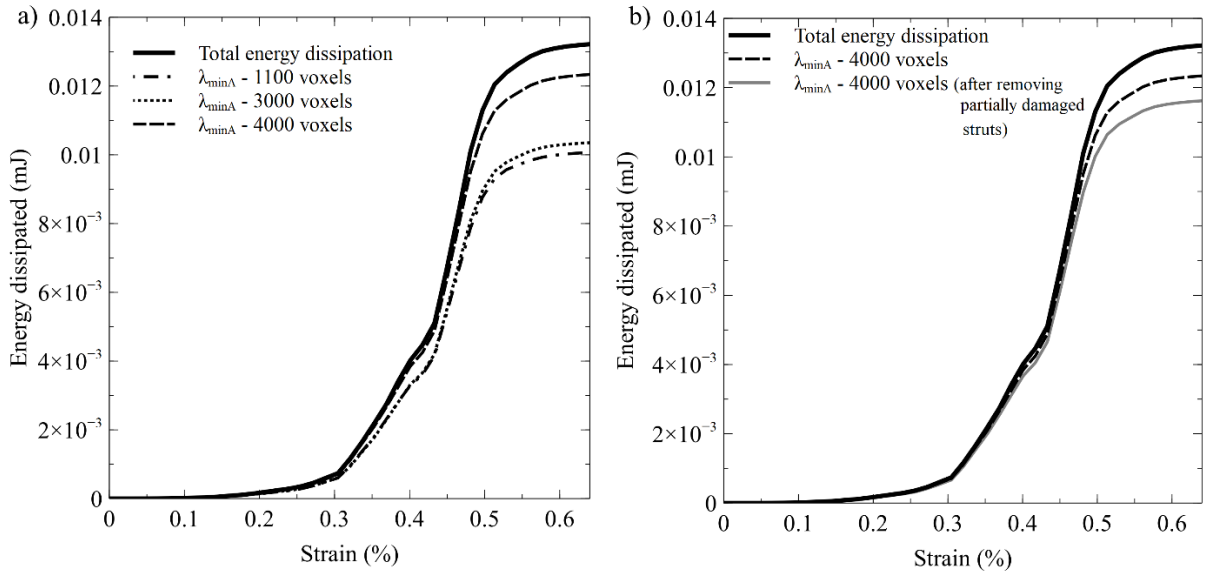

**Fig. 10.** *a) Total dissipated energy by the VE and energy dissipated by the struts for different values of $\lambda_{minA}$; b) total dissipated energy, energy dissipated by the struts for $\lambda_{minA} = 4000$ and energy dissipated after removing partially damaged struts.*

### 3.4 Effectiveness of pruning strategies

Our article [23] showed the effectiveness of the first two types of pruning strategies described in sections 3.2.1 and 3.2.2. Fig.11 shows the number of skeletal branches remaining after applying each pruning strategy along with that of damage and minimum cross-sectional area devised in the present work. After applying all the pruning strategies, only 1763, i.e. about 6 % of the total skeletal branches (27464) are remaining. This means that only 6 % of the skeletal branches in the microstructure are worth studying for the specific problem at hand (uniaxial compression).

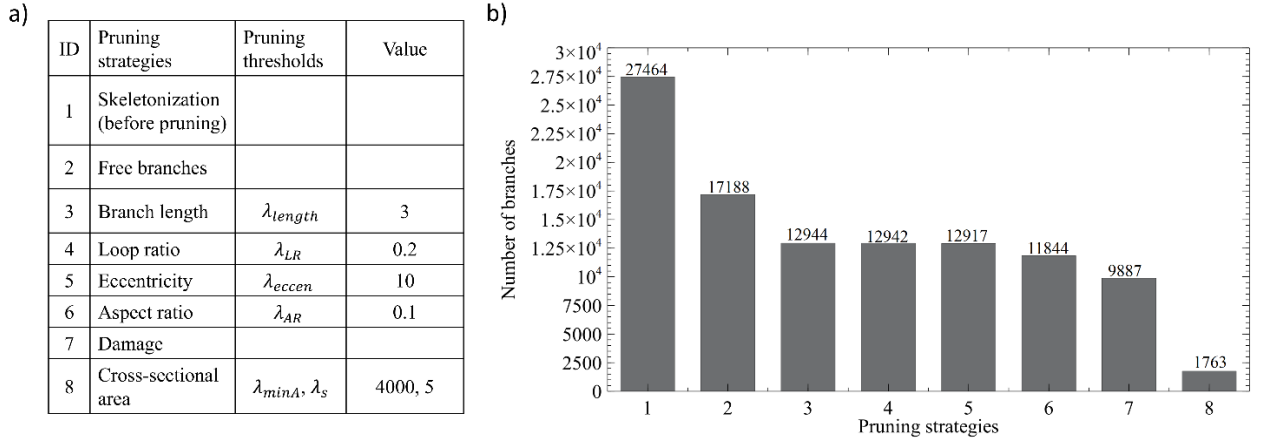

**Fig. 11.** *a) Pruning strategies with IDs and their pruning thresholds; b) number of skeletal branches remaining after every pruning strategy.*

## 4. Geometrical properties of the segmented struts

Orientation of a strut is calculated from the relative position vector defined between the two end points of the skeletal branch. Fig. 12a and Fig. 12b shows orientation of these struts indicated as points on a unit hemisphere. Note that the uniaxial loading on the volume element is vertical, i.e. along X axis (axes are referred in Figs. 12a-b). Visually, it can be seen that the orientation is isotropic in nature. An average second order orientation tensor **N** of all the struts is defined as,

$$\mathbf{N} = \frac{1}{n_t}\sum_{i=1}^{n_t} \mathbf{n}_i \otimes \mathbf{n}_i \ , \quad \text{where} \quad \mathbf{n}_i = \frac{\mathbf{r}_{q_1} - \mathbf{r}_{q_n}}{\|\mathbf{r}_{q_1} - \mathbf{r}_{q_n}\|} \tag{21}$$

and $\mathbf{r}_{q_1}$ and $\mathbf{r}_{q_n}$ are the direction vectors of the end points of the strut skeletal branch $T^{new}(i)$ and $n_t$ is total number of struts. Note that in $\mathbf{n}_i$, i=1,2,3 corresponds to X, Y and Z axes respectively.

$$\mathbf{N} = \begin{bmatrix} \mathbf{0.39} & -0.01 & -0.01 \\ -0.01 & \mathbf{0.33} & -0.00 \\ -0.01 & -0.00 & \mathbf{0.28} \end{bmatrix} \tag{22}$$

The diagonal values of this tensor indicate nearly isotropic behavior. The thickness distribution of the struts is also shown in Fig. 12c.

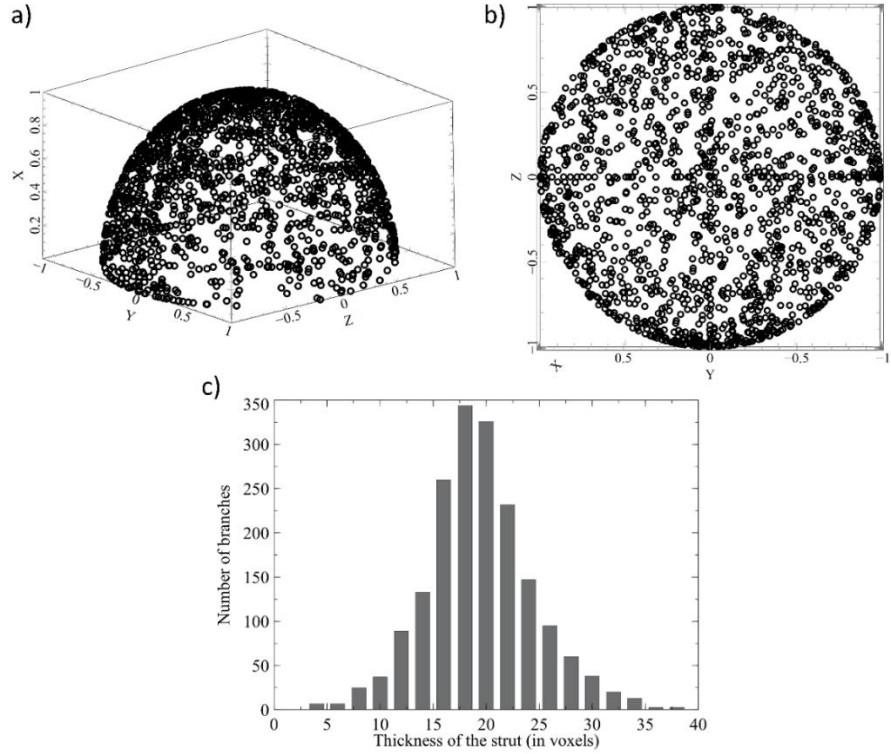

**Fig. 12.** *Orientation distribution of the struts in the VE on a unit hemisphere in a) 3D view (Note that X-axis is the direction of loading) and in b) Y-Z plane; c) thickness distribution of the struts in the VE.*

## 5. Identification of failure modes

### 5.1 Study of damaged and undamaged regions of the strut

In order to study how each of the captured strut fails, it is important to study how each of the finite elements that constitute the strut fail. The strength curve of the ceramic material as defined by the Johnson-Holmquist 2 model takes the form as shown in Fig. 13a. It is a pressure dependent strength model whose compressive strength is much higher than the tensile strength as is the case in typical brittle materials. This difference between the strength in positive and negative pressure states can be utilized to categorize the different failure modes of the struts.

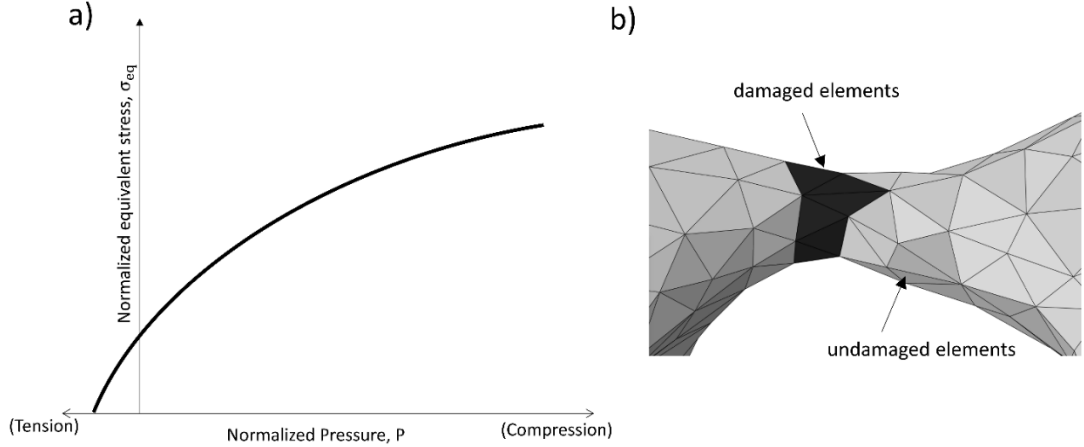

**Fig. 13.** *a) Strength curve of Johnson-Holmquist 2 model; b) a typical damaged strut with damaged elements marked in black color and undamaged (intact) elements marked in grey color.*

The work utilizes a continuum damage model in which each element gets damaged when the equivalent (von Mises) stress in that element reaches the strength curve. The pressure state in each element when the equivalent stress reaches the strength value can be either positive or negative. This can give interesting information about the failure modes. To study this, two quantities have been defined: the volume fraction of the undamaged region of the strut that has positive pressure, $v_{intact}^{+P}$ and the volume fraction of the damaged region of the strut that has positive pressure, $v_{dam}^{+P}$. Fig. 13b shows a typical damaged strut with damaged region marked in black color and undamaged (intact) region marked in grey color.

$$v_{intact}^{+P} = \frac{V_{intact}^{+P}}{V_{intact}} \quad , \quad v_{dam}^{+P} = \frac{V_{damaged}^{+P}}{V_{damaged}}. \tag{23}$$

Note that $V_{intact}$ and $V_{damaged}$ are the intact and damaged volume of the strut respectively. $V_{intact}^{+P}$ and $V_{damaged}^{+P}$ are the intact and the damaged volume respectively of the strut in positive pressure. The elastic energy stored in the strut, $\Psi_{el}$ is also an important information as it shows how much energy is stored in the strut and how it is dissipated as the strut fails. This stored energy is given by Eq. 24 where $\boldsymbol{\varepsilon}^{el}$ is the elastic strain.

$$\Psi_{el} = \int_{V_{E(i)}} \boldsymbol{\sigma}:\boldsymbol{\varepsilon}^{el} \, dV_{E(i)} \tag{24}$$

These three quantities namely, $v_{intact}^{+P}$, $v_{dam}^{+P}$ and $\Psi_{el}$ are calculated for every strut. Fig. 14 shows the relationship between them. The color of the contour represents the order of the maximum stored energy calculated by taking the logarithm of the maximum energy value. The black dots (markers) represent the struts identified by the algorithm. It can be seen that in the domain of Fig. 14, these black dots are not equally spaced with respect to each other. Therefore, the contour plot is obtained by creating another set of points equally spaced from each other and the value of order of energy at these points is calculated by taking the average of black dots in its local circular neighbourhood with radius 0.1. It can be seen that the maximum stored energy increases by orders of magnitude from the bottom left corner to the top right corner.

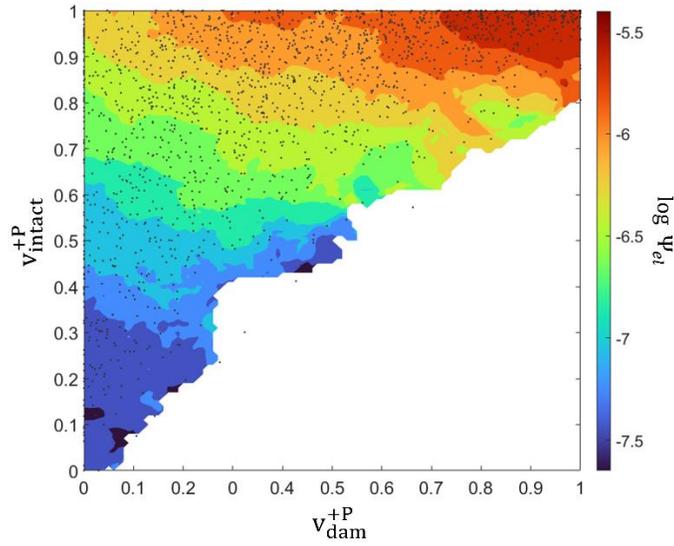

**Fig.14.** *Graph of $v_{intact}^{+P}$ vs $v_{dam}^{+P}$ where the contour color represents $log(\Psi_{el})$.*

Figs.15 shows the relationship between the orientation of the strut and order of energy absorbed. The black color dots indicate the struts and the contour plot is obtained by a procedure similar to the one explained for Fig. 14. It can be seen that struts which absorb low energies (blue color region) have preferred orientation normal to the direction of loading whereas struts which absorb high energies (red color region) have preferred orientation parallel to the direction of loading.

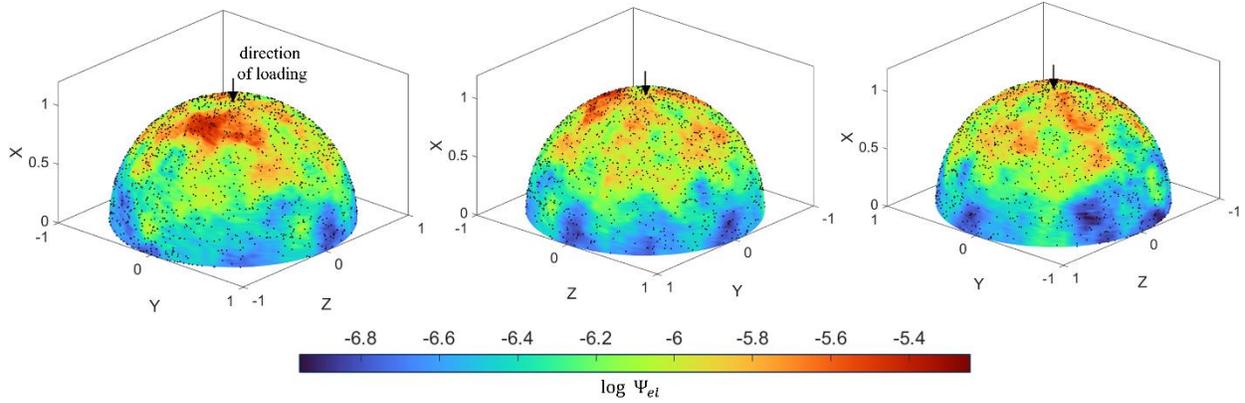

**Fig.15.** *A unit hemisphere showing relationship between strut orientation and $\log(\Psi_{el})$.*

## 5.2 Failure modes

In this section, each strut is individually examined to study its failure mode. The evolution of the quantities $v_{intact}^{+P}$, $v_{dam}^{+P}$ and the elastic energy stored in the strut as the loading increases is studied along with the location of the damaged elements in the strut. Three modes of strut failure are identified as described below.

### 5.2.1 Mode 1 failure

In this type of failure, the major part of the strut experiences negative pressure state as shown in Fig. 16a. The end regions of the strut where it connects to the junction points can experience positive or negative pressure depending on the stress state in the neighbouring struts. The damage in such struts occurs in the central part of strut where the cross-section is at its minimum as shown in Fig. 16b. The evolution of elastic energy absorbed in the strut along with the volume fractions $v_{intact}^{+P}$ and $v_{dam}^{+P}$ as the loading is increased in shown in Fig. 16c. The peak energy value (marked as grey vertical line) corresponds to the state in which the strut is loaded to its maximum (has maximum stored energy) after which it shows loss of strength. Examining the value of $v_{intact}^{+P}$ just before peak energy (dark grey shaded region) shows that its value is very low (around 0.1) which can be seen in Fig. 16a as well since most of the strut is in negative pressure state. The value of $v_{dam}^{+P}$ just after peak energy (light grey shaded region) is 0 since the entire damaged region fails in the state of negative pressure. This type of strut failure is characterized by very low values of $v_{intact}^{+P}$ and $v_{dam}^{+P}$ and the location of damage is in the central region of the strut.

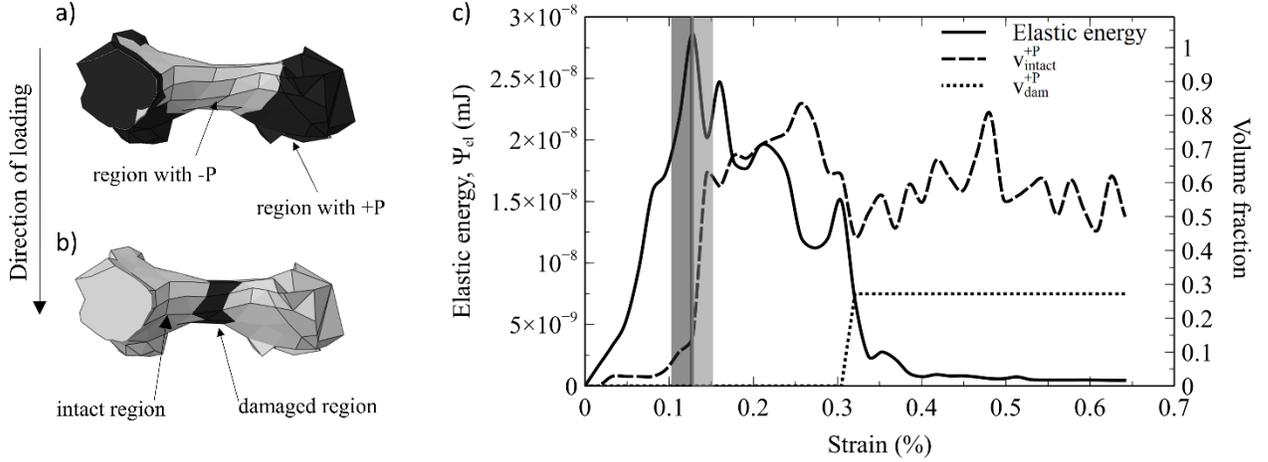

**Fig.16.** *a) Pressure state and b) damaged state in a strut at* $\max(\Psi_{el})$ *that fails in mode 1; c) evolution of* $v_{intact}^{+P}$, $v_{dam}^{+P}$ *and* $\Psi_{el}$ *in a strut with respect to macroscopic strain in the VE.*

### 5.2.2 Mode 2 failure

In this type of failure, the major part of the strut experiences positive pressure state except two small regions near the ends which are on the opposite sides of the strut. Fig. 17a shows pressure state of the strut when it has peak energy. As the loading is increased, these negative pressure regions grow and get damaged since the strength of the base material in tension is much smaller than in compression. Fig. 17b shows damaged state of the strut when it has peak energy. It can be seen in Fig. 17c that the value of $v_{intact}^{+P}$ before peak energy is very high (around 0.95) since most of the strut is in positive pressure state. The value of $v_{dam}^{+P}$ just after peak energy is around 0.5. This indicates 50 % of the damaged region had negative pressure and 50 % had positive pressure.

The pressure values are small in these regions and some failed elements had positive pressure and others had negative pressure. This type of strut failure is characterized by very low value of $v_{intact}^{+P}$ and low to medium value of $v_{dam}^{+P}$ and there are two locations of damage, both near the ends but on the opposite sides of the strut.

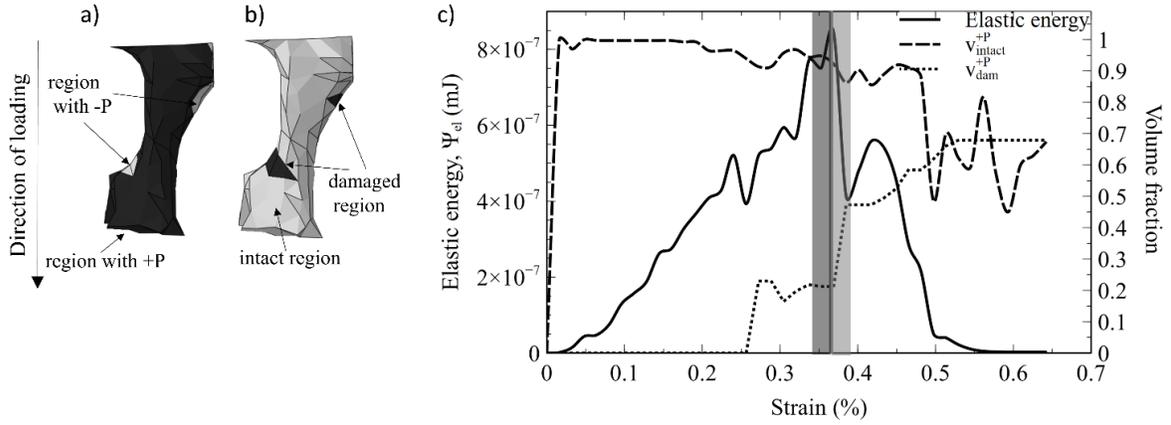

**Fig. 17.** a) *Pressure state and b) damaged state in a strut at* $\max(\Psi_{el})$ *that fails in mode 2; c) evolution of* $v_{intact}^{+P}$, $v_{dam}^{+P}$ *and* $\Psi_{el}$ *in a strut with respect to macroscopic strain in the VE.*

### 5.2.3 Mode 3 failure

In this type of failure, almost the entire strut experiences positive pressure as shown in Fig. 18a. The loss of the strength in such cases is sudden when the central region of the strut which has minimum cross-sectional area fails in compression as shown in Fig.18b. It can be seen in Fig. 18c that the value of $v_{intact}^{+P}$ before peak energy is close to 1 and the value of $v_{dam}^{+P}$ after peak energy is close to 1. This type of strut failure is characterized by very high values of $v_{intact}^{+P}$ and $v_{dam}^{+P}$ and the location of damage is in the central region of the strut.

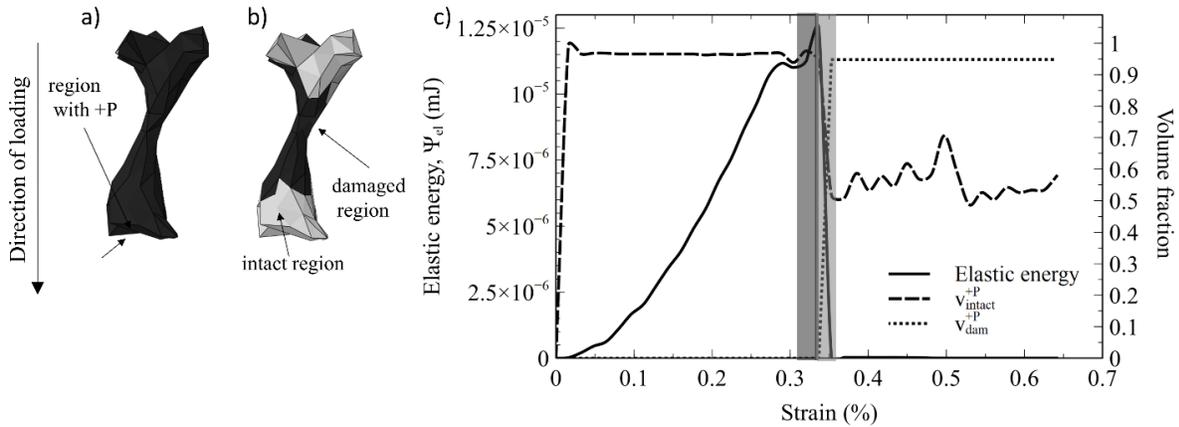

**Fig.18.** a) *Pressure state and b) damaged state in a strut at* $\max(\Psi_{el})$ *that fails in mode 3; c) evolution of* $v_{intact}^{+P}$, $v_{dam}^{+P}$ *and* $\Psi_{el}$ *in a strut with respect to macroscopic strain in the VE.*

Another observation is the maximum amount of energy stored in the struts in all three cases. Since the strength of the base material in tension is very low, the mode 1 struts fail at early stages of the loading. They are not able to absorb a lot of energy. The order of magnitude of energy absorbed is -8. The mode 2 struts are mostly in positive pressure state and hence are able to absorb a lot more energy. The order of energy absorbed is -7. Since the damage occurs in regions of negative pressure the energy absorbed is still much less than mode 3 struts which are entirely in compression and hence are able to absorb energy of the order of -5.

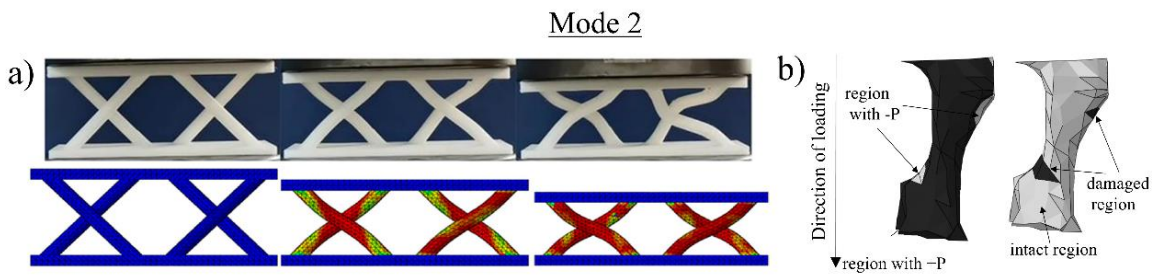

**Fig. 19.** Mode 2 failure observed in a) [24] and b) the present work.

Mode 2 is a mixed mode which has been observed in literature in other applications as well. In [24], 3D printed struts made of polylactic acid (PLA) were manufactured and loaded in uniaxial compression as shown in Fig. 19a. Their simulation results shows red color compression regions in the middle of the struts and green color tension regions at the ends. Yielding also occurs in these tension regions at the ends. In our application, even though the material is brittle, similar stress states are observed (refer Fig. 19b) and the damage occurs in the regions in tension near the ends.

## 5.3 Effect of orientation, $v_{intact}^{+P}$ and $v_{dam}^{+P}$ on types of failure modes

Each of the 1763 struts are studied with respect to the parameters discussed in sections 5.1 and 5.2 and categorized into the three types of failures. The graphs showed in Figs. 14 and 15 are shown again in Fig. 20 with a change that the color of markers now indicate the mode of failure. It can be seen in Fig. 20a that mode 1 struts are concentrated in bottom left of the graph where the values of $v_{intact}^{+P}$ and $v_{dam}^{+P}$ are low. Mode 2 struts are concentrated in top left of the graph where the value of $v_{intact}^{+P}$ is high and the value of $v_{dam}^{+P}$ is low. Mode 3 struts are concentrated in top right corner where the values of both $v_{intact}^{+P}$ and $v_{dam}^{+P}$ are high. The orientation distribution of the three types of struts is shown in Figs. 20b-c. It can be seen that most of the mode 1 struts are oriented normal to the loading direction (X axis) while most of the type 3 struts are oriented parallel to the loading direction. The orientation of mode 2 struts is isotropic.

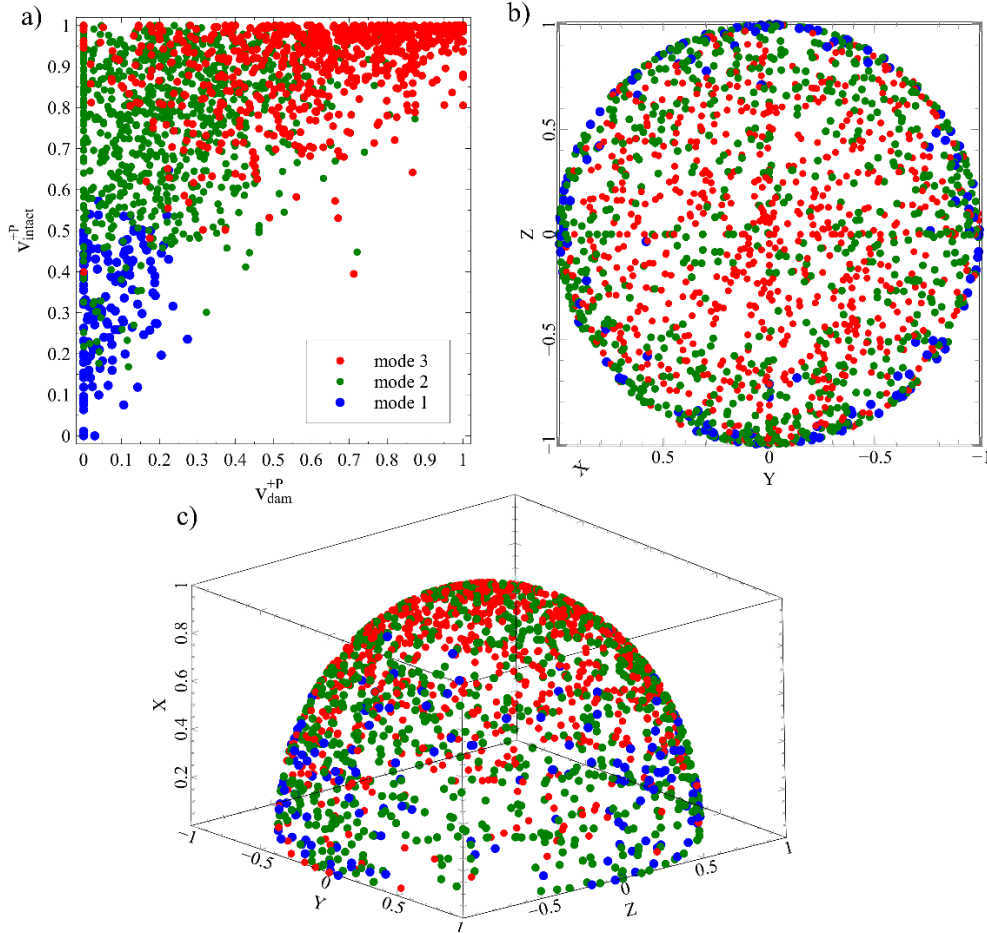

**Fig. 20.** *a) Graph of $v_{intact}^{+P}$ vs $v_{dam}^{+P}$ and orientation distribution of the struts within the VE on a unit hemisphere; in b) Y-Z plane and in c) 3D view where the marker color represents the modes of failure.*

The average second order orientation tensors for the three types of struts are shown below.

$$\mathbf{N}^{mode1} = \begin{bmatrix} \mathbf{0.07} & -0.01 & -0.01 \\ -0.01 & \mathbf{0.51} & 0.03 \\ -0.01 & 0.03 & \mathbf{0.41} \end{bmatrix} \tag{25}$$

$$\mathbf{N}^{mode2} = \begin{bmatrix} \mathbf{0.31} & -0.01 & -0.01 \\ -0.01 & \mathbf{0.37} & -0.01 \\ -0.01 & -0.01 & \mathbf{0.32} \end{bmatrix} \tag{26}$$

$$\mathbf{N^{mode3}} = \begin{bmatrix} \mathbf{0.53} & -0.01 & 0.00 \\ -0.01 & \mathbf{0.24} & 0.00 \\ 0.00 & 0.00 & \mathbf{0.22} \end{bmatrix} \tag{27}$$

The diagonal terms of the tensors reinforce the earlier statements. In case of mode 1 struts, $N_{11}^{mode1} \ll N_{22}^{mode1}$, $N_{33}^{mode1}$ and $N_{22}^{mode1} \approx N_{33}^{mode1}$ which indicates the orientation of struts is isotropic in the plane normal to the loading direction. In case of mode 2 struts, $N_{11}^{mode2} \approx N_{22}^{mode2} \approx N_{33}^{mode2}$ which indicates the orientation is isotropic. In case of mode 3 struts, $N_{11}^{mode3} > N_{22}^{mode3}$, $N_{33}^{mode3}$ and $N_{22}^{mode3} \approx N_{33}^{mode3}$ which indicates preferred orientation as parallel to the loading axis.

## 6. Calculating the energy dissipated by the different failure modes

In order to understand the relative impact of the different failure modes on the overall failure of the VE, the energy dissipated $\Psi_d$ by the struts belonging to the three different categories of failure is calculated. Fig. 21a shows the energy dissipated by the three types of struts at each load step.

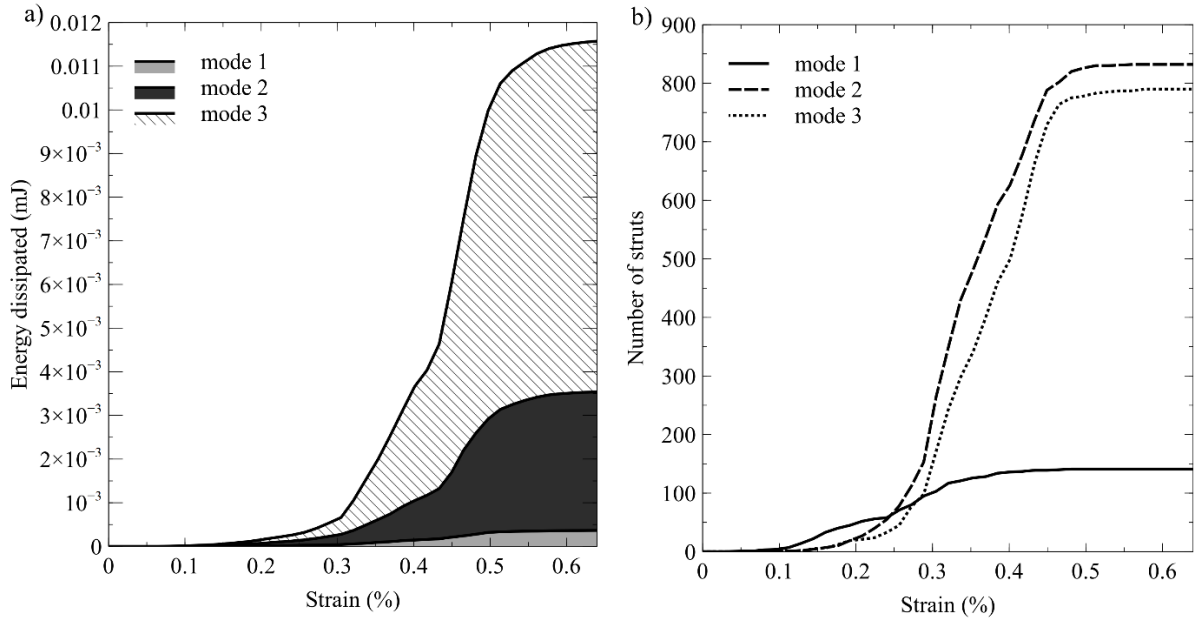

**Fig.21.** *a) Energy dissipated by the three types of struts based on failure modes; b) number of struts failed in each mode at each load step.*

The shaded regions indicate the dissipated energy. It can be seen that the energy dissipated by mode 1 struts is negligible as compared to the other two types. Mode 2 struts dissipate considerable energy but Mode 3 struts dissipate the highest amount of energy. Note that addition of energies dissipated by all three modes equals the total energy dissipated by the VE as shown in Fig. 10b. Fig. 21b shows the number of struts failed in three failure modes at each load step. The load step of failure corresponds to the step at which the energy absorbed by the strut is maximum. It can be seen that as the loading is increased, struts start failing in mode 1 failure. Beyond 0.3% strain, the struts do not fail in mode 1 but in modes 2 and 3. This strain value also corresponds to the point at which the macroscopic stress in the VE is at its maximum (refer Fig. 3a).

## 7. Discussion

This work studies the uniaxial compression failure of a ceramic foam manufactured by direct foaming method. The foam microstructure is characterised by homogeneous distribution of spherical pores with a wide distribution in their sizes. The work focuses on studying how the struts in the microstructure fail and how it affects the overall failure of the material.

A large body of literature e.g. [7, 8, 21] show that uniaxial compression failure of brittle foams occurs when the struts in the microstructure fail in bending. Since the tensile strength of most brittle materials is much smaller than that in compression, the crack initiates at the outer surface of the strut that is in tension and then it moves inwards leading to complete rupture of the strut. However, when the same loading was applied on the particular material studied in this work, the effective stress-strain curve obtained (Fig. 3a) through experimental measurements [14] as well as simulations [15] was much different than the typical curve (Fig. 4d). The strength of the foam material was also much higher than the one predicted by Gibson-Ashby model [21].

To understand this difference, the first task is to study the geometry of the struts in the material microstructure. This is an image segmentation task in which the binary digital image of the foam microstructure is to be segmented so as to isolate the struts and study their properties. In our recent article [23], a skeletonization based image segmentation algorithm was developed in which a series of novel pruning strategies segment the struts in the microstructure on the basis of their geometry. In the studied problem of uniaxial compression, the macroscopic failure occurs due to failure of the struts. Therefore, the only requirement on the pruning strategies is to not prune any failed strut. The present work introduces a pruning strategy based on the damage in the struts which by definition satisfies this criterion. Note that this pruning strategy is problem specific in the sense that it prunes many geometrically well-defined struts just because they are not damaged under the considered loading. The most important pruning strategy proposed in [23] is that of the minimum cross-sectional area because the value of its threshold parameter decides which struts in the microstructure will be segmented. All the struts that have minimum cross-sectional area below the threshold value are captured by the algorithm. As the threshold value is increased, a greater number of struts are captured. So, in this case a proper tuning strategy is necessary. Since the objective of the present work is to identify the struts that caused macroscopic failure, the total energy dissipated during the loading acts as a reference for tuning. For different values of $\lambda_{minA}$, different number of struts are stored and energy dissipated by these struts during loading is calculated. This is compared by the total energy dissipated by the microstructure. For $\lambda_{minA}$=4000, almost all the dissipated energy of the microstructure was captured by the struts as shown in Fig. 10a. Hence this value is decided as the threshold value. For studying any other physical problem, physics-based measures relevant to that problem should be chosen. It can be seen in Fig. 11b that pruning strategies of free branches and cross-sectional area are particularly useful as they prune a significant number of skeletal branches. The damage-based pruning measure is also effective in pruning a lot of branches not relevant to the studied problem.

The struts that remained (not pruned) after applying all the pruning strategies are studied in terms of their thickness and orientation distribution within the microstructure. Figs.12a-b as well as Eq.22 show that the orientation distribution is isotopic.

The failure of each of the remaining struts is studied by observing the state of stresses obtained from the finite element simulations. The volume fractions $v_{intact}^{+P}$ and $v_{dam}^{+P}$ define the relative undamaged and damaged regions of the strut respectively that are in positive pressure state. This distinction between the positive and negative pressure state is important because the strength of the base material in negative pressure (tension) is much smaller than that in positive pressure (compression). The total elastic energy $\Psi_{el}$ absorbed by each strut during loading is also calculated whose peak value gives an indication of how much energy the strut absorbs and when it losses its strength. Fig. 14 shows that there is relationship between $v_{intact}^{+P}$, $v_{dam}^{+P}$ and the order of energy absorbed by each strut. There is an increase in the order of energy as $v_{intact}^{+P}$ and $v_{dam}^{+P}$ increase. Fig. 15 shows that there is a correlation between orientation of strut and the energy absorbed. Low energy struts (blue region) are concentrated near the equator of the hemisphere indicating that they have preferred orientation normal to the direction of loading. High energy struts (red region) are concentrated near the pole of the hemisphere indicating that they have preferred orientation parallel to the direction of loading. The different colors in the contour do not change smoothly with respect to the orientation but are distributed in patches. This is due to lack of sufficient number of struts and lack of homogeneous distribution of struts in the orientation space. More uniform variation of energy could be obtained by increasing the number of struts which requires studying bigger microstructure sample. But this is computationally very expensive.

The identification of modes of failure of the struts is also done by studying the evolution of $v_{intact}^{+P}$, $v_{dam}^{+P}$ and $\Psi_{el}$. In mode 1, the strut failure occurs due to tension or negative pressure state in the strut. This type of failure is similar to the stretching failure reported in the literature [7, 8, 10, 11, 12]. In the present microstructure, some struts have pressure distribution that resembles a bending deformation. But due to the complicated geometries of the struts it was difficult to distinguish between bending and stretching failures. Hence, they were categorized together as mode 1 failure. Mode 2 failure is interesting because even though most of the strut region is in positive pressure state, the failure occurs in the two small regions of negative pressure near the ends. In mode 3 failure, almost the entire strut is in positive pressure state and hence it fails in the same state. This is a typical uniaxial compression failure of the strut. Most of the struts have very small aspect ratios (refer [23]) to cause any other type of failure like buckling. Another distinguishing factor between the three failure modes is the location of the failure. In mode 1, the failure occurs at the center of the strut where the cross-section is at its minimum. In mode 2 the failure occurs at the end regions and on opposite sides of the strut. In mode 3 the location is again in the center where the cross-section is at its minimum. There is order of magnitude difference between the energies absorbed by the three types of struts as shown in Figs.16c, 17c, 18c. This has a direct effect on the relative impact of the failure modes on the overall failure of the VE.

These observations are used to categorize the struts into the three different types. Fig. 20a shows 3 distinct regions. Mode 1 with low values of $v_{intact}^{+P}$ and $v_{dam}^{+P}$, mode 2 with low value of $v_{intact}^{+P}$ and low to medium value of $v_{dam}^{+P}$ and mode 3 with high values of $v_{intact}^{+P}$ and $v_{dam}^{+P}$. There are certainly outliers to this categorization. This is mainly because of the approximations utilized in defining the volume occupied by the strut and the finite element discretization scheme. Better resolution of the digital image and even finer discretization might lead to better categorization. Figs. 20b-c and the orientation tensors in Eqs. 25-27 show that there is indeed a relation between failure modes 1 and 3 with the orientation of the struts. Most struts in mode 1 failure are oriented normal to the loading direction. Most struts in mode 3 failure are oriented parallel to the loading direction. Struts failing in mode 2 do not have any preferred orientation.

The dominance of failure modes of the struts on the overall failure of the VE is studied by calculating the energy dissipated by the struts belonging to each failure mode. Fig. 21a shows that the most dominant mode is mode 3 followed by mode 2 whereas the effect of mode 1 is negligible. Fig. 21b shows that mode 1 struts start failing at around 0.07% strain and the number keeps rising steadily. However, the macroscopic stress in the VE keeps on increasing (refer Fig. 3a) as mode 1 is not a dominant mode of failure. The struts in mode 2 and 3 types of failure start failing at around 0.14 % strain but the number start increasing rapidly at around 0.2 % strain for mode 2 and 0.25 % strain for mode 3. After 0.3 % strain, the number of struts failing in mode 1 is negligible and most of the struts fail in modes 2 and 3. The VE has peak stress at 0.3 % strain and it decreases after that. Even the plastic energy dissipated by modes 2 and 3 rapidly increases after 0.3 % strain which indicates that the overall failure of the VE occurs because of modes 2 and 3 rather than mode 1. This shows that the dominant failure modes in this type of microstructure are much different than that reported in the literature. The reason is mostly attributed to the presence of low aspect ratio struts in the microstructure that fail in compression. The mode 2 failure is more complicated and probably results from the complex loading pattern on these struts.

## 8. Conclusion

The questions raised in the introduction are answered as follows. The theoretical model of Gibson-Ashby does not work in the case of studied microstructure because the dominant failure modes are different than the bending failure as proposed in the GA model. The most dominant failure mode is compression followed by a mixed mode in which most of the strut has positive pressure state but the failure occurs in negative pressure regions at the ends. The effect of stretching/bending failure mode which is fundamentally similar to the GA model is negligible. As a result of this, the effective stress-strain curves are also much different than the GA model. During the early stages of the loading when most of the struts are failing in mode 1, the other struts are taking the load and hence the macroscopic stress keeps increasing. After peak stress, there is a sudden increase in the number of struts failing in mode 2 and 3 whereas there is negligible increase in struts failing in mode 1. So, the loss of strength occurs due to modes 2 and 3 rather than mode 1. Studying these two dominant failure modes is important to devise any theory applicable to such material microstructures.

**Acknowledgements**

The financial support of the University of Applied Sciences Darmstadt is gratefully acknowledged. This research was supported by the Hessian Ministry of Higher Education, Research, Science and the Arts, Germany within the Framework of the "Programm zum Aufbau eines akademischen Mittelbaus an hessischen Hochschulen''. This research was also funded by the Federal Ministry for Economic Affairs and Energy based on a decision of the German Bundestag.

**Declaration of competing interest**

The authors declare that they have no known competing financial interests or personal relationships that could have appeared to influence the work reported in this paper.